\documentclass[submission,copyright,creativecommons]{eptcs}
\providecommand{\event}{HAS 2014} % Name of the event you are submitting to 

\usepackage{amsmath}
\usepackage{amssymb}
\usepackage{graphicx}
\usepackage{cases}
\usepackage[tight,footnotesize]{subfigure}
\usepackage{hyperref}

\usepackage[crop=pdfcrop]{pstool}

\usepackage{color}
\usepackage[numbers]{natbib}     % required for bibliography

\usepackage{url}
\urldef{\mailsc}\path|{lct,raf,jjl}@es.aau.dk|

\providecommand{\urlalt}[2]{\href{#1}{#2}} 
\providecommand{\doi}[1]{doi:\urlalt{http://dx.doi.org/#1}{#1}}

\title{Modeling Populations of Thermostatic Loads with Switching Rate Actuation\thanks{This work is supported by the Southern Denmark Growth Forum and the
European Regional Development Fund under the project ``Smart \& Cool".}}

\author{Luminita C. Totu \qquad Rafael Wisniewski \qquad John Leth
\institute{Automation and Control, Aalborg University, Denmark}
\email{lct,raf,jjl@es.aau.dk}
}
\def\titlerunning{Modeling Populations of Thermostatic Loads with Switching Rate Actuation}
\def\authorrunning{L. C. Totu, R. Wisniewski \& J. Leth}

\begin{document}
\maketitle

\begin{abstract}
We model thermostatic devices using a stochastic hybrid description, and introduce an external actuation mechanism that creates random switches in the discrete dynamics. 
We then conjecture the form of the Fokker-Planck equation and successfully verify it numerically using Monte Carlo simulations. 
The actuation mechanism and subsequent modeling result are relevant for power system operation.
%\textit{Markov process, hybrid system, Fokker-Planck equation, thermostatic load, demand response}
\end{abstract}

\section{Introduction}

In the context of power system operation and Smart Grids technologies, thermostatically controlled loads (TCLs), such as refrigerators, air-conditioners or heat-pumps, are seen as a promising resource of demand response services \cite{IREP2013,CAL2}. Essentially, TCLs have the potential of acting as distributed energy storages that can be scheduled and controlled to balance out grid fluctuations. Arguably,  this can be used to decrease the overall capacity requirements for spinning reserves, and contribute towards the integration of more intermittent generation, such as wind, into the grid. 

Since the individual TCL has a very small energy storage capacity relative to the scale of power system operation, any relevant demand response strategy requires the participation of a very large number of TCLs. For this reason, developing demand response algorithms requires not only models for individual TCLs, but also models for TCL populations. An overview of recent population modeling results can be found in \cite{MO2013}. 

This work presents an aggregate model for a TCL population under a specific demand response strategy, the Switching Rate broadcast actuation. This actuation is closely related to the Switching Fraction broadcast proposed and analyzed in \cite{zhang2013aggregated,totu2013control,totu2014demand}, but has the added advantage that the switching actions are not synchronized across the population. An individual TCL unit is modeled as a Stochastic Hybrid System (SHS) with the Markov property, and the resulting population model is in the form of a Partial Differential Equation (PDE) or Partial Integro-Differential Equation (PIDE) system and boundary conditions. This PDE form corresponds to a generalized Fokker-Planck (Forward Kolmogorov) operator \cite{bect2010unifying} associated with the TCL stochastic hybrid system. 

The Fokker-Planck approach for TCL population modeling is not in itself new. It was first used in \cite{malhame1985electric} for modeling a TCL population without (continuous) external actuation. However, to the best knowledge of the authors, the Switching Rate actuation variant and the resulting population model are new and should be a useful contribution to the topic. 
 
The article is organized as follows. The stochastic hybrid model of the TCL unit and the Switching Rate actuation are presented in Section \ref{sec:SHS}. PDE population models are then given in Section \ref{sec:pde_models}.  Numerical simulations and results addressed in Section \ref{sec:numerical_sim}, while Section \ref{sec:conclusion} points to future work.

\section{Stochastic Hybrid Model for the TCL Unit} \label{sec:SHS}

Similar to other works, we consider that the TCL unit can be abstracted as a hybrid dynamical system with temperature as a continuous state and the power mode, ``on" or ``off", as a discrete state. An informal presentation follows next where mathematical constructions are not rigorously addressed, but remarks about the formal setting are made towards the end of the section. 

\subsection{Unactuated TCL}

We consider the dynamics of the continuous state represented by Stochastic Differential Equations (SDE) of the following form,
\begin{subequations} \label{eq:cont_dyn}
\begin{align} 
 dT_t  &= u_0(T_t,t)dt +  \sigma_0(T_t,t) dw_t, \hspace{10pt} \text{for } m_t=0 \\
 dT_t  &= u_1(T_t,t)dt +  \sigma_1(T_t,t) dw_t, \hspace{10pt} \text{for } m_t=1  ,
\end{align}
\end{subequations} 
where $T \in \mathbb{R}$ is the temperature state, $u_0(\cdot), u_1(\cdot): \mathbb{R}\times[0,\infty) \rightarrow \mathbb{R}$ are deterministic, (potentially) time-varying vector fields, $w_t \in \mathbb{R}$ is a Wiener process, $\sigma_0(\cdot),\sigma_1(\cdot): \mathbb{R}\times[0,\infty) \rightarrow \mathbb{R}^{n\times m}$ are diffusion coefficients, and $m_t \in \{0,1\}$ is the mode state. We chose one-dimensional spaces for the continuous state $T_t$ and the Brownian motion $w_t$ since it simplifies presentation, but other low-dimensional spaces  (e.g. \cite{zhang2013aggregated} uses a two-dimensional temperature state)  could also be considered and the subsequent population model carries over in a straightforward manner. However, it is noted that numerical analysis becomes more difficult as the state space increases, as the Fokker-Planck approach suffers from the curse of dimensionality and can become intractable.

The dynamics of the discrete state involve a thermostat mechanism that is considered equivalent to a state dependent, deterministic rule. For example, in the case of a cooling unit, this can be described as
\begin{align}
  m_{t^+} = \begin{cases} 
                       0, \hspace{5pt} \text{ if } T_{t} \leq T_{\mathrm{min}}  \text{ and } \hspace{3pt} m_{t^-} = 1 \\
                       1, \hspace{5pt} \text{ if } T_{t} \geq T_{\mathrm{max}} \text{ and } \hspace{3pt} m_{t^-} = 0 \\
                      m_{t^-}, \hspace{5pt} \text{otherwise}
                 \end{cases} ~, \label{eq:discrete_dyn}  
\end{align}
where function argument notations $t^+$ and $t^-$ denote limit from the right and from the left respectively, and $T_{\min}$ and $T_{\max}$ are the thermostat boundaries. %The thermostat boundary values could be considered time-varying in a more general case. 
In the multidimensional case, the thermostat-triggering temperature has to be one of the states.
% \footnote{Because $T_t$ is has an a.s. continuous realization and $m_t$ is  piecewise continuous, these limits exist.}

The output of the TCL unit is represented by the instantaneous power consumption $y_t \in \mathbb{R}_{+}$, which must be a function of at least $m_t$. More specific, we consider that the power consumption is constant $r>0$ if the mode is ``on" and is zero otherwise, 
\begin{align}
    y_t = r m_t, \text{  } r>0 \text{  .} \label{eq:output_eq}
\end{align}

\subsection{Switching Rate Actuation}
To make demand response possible, a control element needs to be introduced. The objective is to create the possibility of modifying the power consumption pattern of the TCL unit (and thus also that of the population) in a non-disruptive manner, from an external channel. Non-disruptive means that the TCL temperature is maintained within the thermostat dead-band at all times and no other operational constraints are broken. The Switching Rate mechanism achieves this objective by adding a control element to the discrete-state dynamics.

The idea is to introduce, in addition to the thermostat, a new type of switching: rate-switching. While the thermostat switching is governed by a deterministic law, rate-switching will take place according to a probabilistic law parameterized by an external signal. The external signal will control the rate of occurrence of the probabilistic switches (the average number of switching events in a given period of time). 
%to control the thermal energy storage and release. 
%The exponential distribution is chosen because it is the only distribution that has the memoryless property and can preserve the Markov property of the overall SHS. 

Furthermore, a practical consideration needs to be addressed. A frequent switching behavior is undesirable because it can damage the equipment (e.g. compressor components) and because it is inefficient. If the cooling/heating cycle is active for only a short period of time, it will not produce any significant temperature effect. In addition, temperature dynamics of type \eqref{eq:cont_dyn} will be highly inexact in such cases. 

To avoid frequent switching, two heuristics are added. First, we will prevent the pattern of a thermostat- and a rate-switch occurring closely in time. This is done by allowing rate-switching only if the temperature is a safe distance away from the relevant thermostat boundary. For example, in the case of a cooling unit, switch-off actions are allowed only when the temperature is some distance away from the upper bound (hot zone) of the thermostat interval, and similarly, switch-on actions are allowed only if the temperature is some distance away from the lower bound (cold zone). In this way,  thermostat- and rate-switches will not compete. Second, we will prevent multiple rate-switches to occur closely in time. This is done by imposing a minimum dwell time for modes ``on" and ``off".

The Switching Rate mechanism is described next using more mathematical terms, but the presentation remains informal.  

We introduce $\Delta T_0$ and $\Delta T_1$ as the safe distances from the thermostat boundaries, and add a new continuous state $d_t \in \mathbb{R_+}$, the dwell time. The dwell time acts as a clock variable, $\dot{d}_t = 1$, and resets to zero after each switch. We denote by $M_0$ the minimum dwell time in the  ``off" state, and by $M_1$ the minimum dwell time in the ``on" state. The external control signals for the switch-off and switch-on rates are $\epsilon^0_t$ and $\epsilon^1_t$ respectively. The probability of a rate-switching event in a small time interval $\tau<<1$ can be described as,
\begin{subequations}\label{eq:rate_law}
\begin{align}
&\mathrm{Pr}\big[~ m_{t+\tau} = 1 ~\big|~ m_t = 0 \wedge  T_t \in [ T_{\min} +\Delta  T_1, T_{\max}) \wedge d_t \geq M_0 \wedge \epsilon^1_t~\big]=\nonumber \\
&\hspace{240pt} = \lambda_1(\epsilon^{1}_t,T_t) \tau + o(\tau) ~, \label{eq:rate_on_law} \\ % \hspace{30pt}=  1 - e^{\int_t^{t+\tau} \lambda_1(\epsilon^{1}_s,T_s) ds}  \\
&\mathrm{Pr}\big[~ m_{t+\tau} = 0 ~\big|~ m_t = 1 \wedge  T_t \in ( T_{\min}, T_{\max}- \Delta T_{0}] \wedge d_t \geq M_1 \wedge \epsilon^0_t ~\big]= \nonumber \\
&\hspace{240pt}  = \lambda_0(\epsilon^{0}_t,T_t) \tau + o(\tau) ~, \label{eq:rate_off_law} % \hspace{30pt}=  1- e^{\int_t^{t+\tau} \lambda_0(\epsilon^{1}_s,T_s) ds} 
\end{align}
\end{subequations} 
where the temperature ranges in the conditional part of the probability are exemplified for a cooling unit, and $\lambda_1$ and $\lambda_0$ are real and positive valued rate-functions, which can be seen as part of the (control) design. A straightforward and simple choice for these functions is a temperature-independent form,
\begin{align}
	 \lambda_i(\epsilon,T)=\epsilon, ~ i\in \{0,1\}~. \label{eq:simple_rate_functions}
\end{align}
%Another way to describe the rate-switch mechanism of a TCL unit is to define a stochastic process $N_t$ that counts of the number of probabilistic switches until time $t$. In this case, we can write
%\begin{subnumcases}{ \label{eq:count_law}  }
%\mathrm{Pr}\big[ N_{t+\Delta t} - N_{t} = k ~\big|~ m_t = 0 \wedge T_s \in [ T_{\min} +\Delta  T_1, T_{\max}) \wedge d_t \geq M_0  \wedge \epsilon^1_s ~\big]\nonumber \\
%\hspace{10pt} =  \left( \int_t^{t+\Delta t} \lambda_1(\epsilon^{1}_s,T_s) ds \right)^k e^{ - \left( \int_t^{t+\Delta t} \lambda_1(\epsilon^{1}_s,T_s) ds \right) }, ~t \leq s \leq t+\Delta t \label{eq:count_law_1} \\ 
%\mathrm{Pr}\big[ N_{t+\Delta t} - N_{t} = k ~\big|~ m_t = 1 \wedge T_s \in ( T_{\min}, T_{\max}- \Delta T_0 ] \wedge d_t \geq M_1 \wedge \epsilon^0_s ~\big] \nonumber \\
%\hspace{10pt}  = \left( \int_t^{t+\Delta t} \lambda_0(\epsilon^{1}_s,T_s) ds \right)^k e^{ - \left( \int_t^{t+\Delta t} \lambda_0(\epsilon^{1}_s,T_s) ds \right) }, ~t \leq s \leq t+\Delta t \label{eq:count_law_2} 
%\end{subnumcases}
%where $k \in \{0,1\}$ ($k<2$ because multiple switch-on events cannot happen consecutively, same for switch-off). The above conditional probabilities are similar to a non-homogeneous Poisson process. Description  \eqref{eq:count_law} is equivalent to \eqref{eq:rate_law} as $\Delta T \rightarrow 0 $.

Compared to the Switching Fraction approach \cite{zhang2013aggregated,totu2013control,totu2014demand}, the Switching Rate actuation has the advantage that individual switch events will not be synchronized across the population.
This is useful for at least one reason. It is well known that the power consumption of an individual TCL exhibits a peak (compressor peak) right after switch-on and before converging to the nominal value. This is not captured in the modeling \eqref{eq:output_eq}, and could in practice cause short but high demand peaks that negatively impact grid stability if the switch-on actions are synchronized. %With a Switching Fraction actuation, individual switch-on are nice

\subsection{Remarks on a GSHS description}

The TCL unit could formally be described in the framework of Generalized Stochastic Hybrid Systems (GSHS) \cite{bujorianu2006toward}.
A GSHS is a hybrid system where the continuous states evolve according to a SDEs (as is the case of \eqref{eq:cont_dyn}), 
and where the discrete dynamics can produce jumps in the continuous state (as it the case with the reset of the dwell time state $d_t$).
Furthermore, the discrete dynamics are described by probabilities (in the TCL case, the switch-rate laws \eqref{eq:rate_law}), 
or occur when the continuous state hits a certain domain boundary (in the TCL case, the thermostat mechanism). 
A GSHS has the strong Markov property and trajectories that are right continuous with left limits.

The only issue that needs to be addressed is the fact that the GSHS definition does not explicitly include dependences of an external control element, 
as is the case of the transition rate functions, or time, as is the case with the continuous dynamics. We postpone this technical discussion for future work.

\section{Probability Density Model}  \label{sec:pde_models}

In the absence of the Switching Rate mechanism, the TCL unit can be described, equivalent in effect with the SHS characterization,
in terms of the probability density function (pdf) over the hybrid state space $(T,m) \in \mathbb{R}\times \{0,1\}$, namely
\begin{align}
     f_i(x,t) = \frac{1}{ dx} \mathrm{Pr}\big[  \hspace{3pt}T(t) \in ( x, x + dx] \hspace{3pt} \wedge \hspace{3pt} m(t)=i  \hspace{3pt}~\big]~. \label{def:pdf}
\end{align} 
Building on elements and results from Markov process theory (e.g. \cite{dynkin1965markov,gardiner1985handbook}), \cite{malhame1985electric} showed that the dynamic of $f_i(x,t)$ can be described analytically. In particular, the dynamic of  $f_i(x,t)$ represents the generator of the forward-operator linear semigroup associated with the TCL  SHS. For dynamical systems characterized by regular SDEs, without hybrid elements, this generator is known as the Fokker-Planck equation. Therefore, the result in \cite{malhame1985electric} can be seen as a Fokker-Planck operator specific to the TCL SHS.

Unlike the SHS form, a TCL description in terms of the pdf translates almost directly into a (homogeneous) population model.  Probability quantities simply change meaning to  population fractions, see e.g. \cite{malhame1985electric} and \cite{totu2014demand}. The latter contains also a discussion and results on heterogeneous populations.

\subsection{Unactuated TCL}

For an unactuated TCL unit, \cite{malhame1985electric} showed that the dynamics of $f_i(x,t)$ can be described by a system of Fokker-Planck equations, each acting on a sub-domain of the hybrid state-space. These sub-domains are $0a= (-\infty, T_{\min}) \times \{0 \}$, $0b= (T_{\min}, T_{\max}) \times \{0\}$, $1b= (T_{\min}, T_{\max}) \times \{1\}$, $1c= (T_{\max}, \infty) \times \{1\}$, and the pdf $f_i(x,t)$ is reconstructed from four segments, $f_{0a}$, $f_{0b}$, $f_{1b}$, and $f_{1c}$. The separation of the pdf into components $f_0$ and $f_1$ corresponds to the ``off" and ``on" discrete modes, and it appears naturally as seen already in \eqref{def:pdf}. The partition of the temperature domain into the regions $a$, $b$ and $c$, is a result of the pdf $f_i(x,t)$ not being $x$-differentiable at $T_{\min}$ and $T_{\max}$. Furthermore, the pdf is zero over the omitted domains $1a$ and $0c$. These features are a result of the thermostat. 
The dynamic for each pdf segment is given by the Fokker-Planck equations
\begin{align}
   \frac{\partial f_{ip}}{\partial t}(x,t) = - \frac{\partial}{\partial x} \bigg( u_i(x,t) f_{ip}(x,t) \bigg) +  \frac{\partial^2}{\partial x^2} \left( \frac{1}{2} \sigma_i^2(x,t) f_{ip}(x,t) \right), \label{eq:FP}
\end{align}
and the following boundary conditions apply,
\begin{subequations}   \label{eq:PDE_boundary} 
\begin{align}
  & f_{1b}(T_\mathrm{min},t) = 0,~  f_{0b}(T_\mathrm{max},t) = 0  \\
  &f_{0a}(-\infty,t) = 0, ~ f_{1c}(+\infty,t) = 0  \\
  &f_{0b}(T_\mathrm{min},t) =  f_{0a}(T_\mathrm{min},t),~  f_{1b}(T_\mathrm{max},t) =  f_{1c}(T_\mathrm{max},t) \\
  & h_{0a}(T_\mathrm{min},t) =  h_{0b}(T_\mathrm{min},t) + h_{1b}(T_\mathrm{min},t) \label{eq:B5}\\ 
  & h_{1c}(T_\mathrm{max},t) =  h_{0b}(T_\mathrm{max},t) + h_{1b}(T_\mathrm{max},t) \label{eq:B6} 
\end{align}
\end{subequations}
where $i \in \{0,1\}$ and $p \in \{a,b,c\}$, in the allowed combinations mentioned above, $h_{ip}(x,t)$ are probability flows defined as $\int \frac{\partial f_{ip}}{\partial t} dx$,  and  \eqref{eq:B5} and \eqref{eq:B6} are particular to the case of a cooling unit. For the differential forms in the right hand side of \eqref{eq:FP} to exist, it is implied that the functions $u_i$ and $\sigma_i$ need to be sufficiently smooth. 
%  &\frac{\partial}{\partial x} f_{1b}(T_\mathrm{min},t) +  \frac{\partial}{\partial x} f_{0b}(T_\mathrm{min},t) - \frac{\partial}{\partial x} f_{0a}(T_\mathrm{min},t)  =0  \label{eq:B5}\\
%  &\frac{\partial}{\partial x} f_{1b}(T_\mathrm{max},t) +  \frac{\partial}{\partial x} f_{0b}(T_\mathrm{max},t) - \frac{\partial}{\partial x} f_{1c}(T_\mathrm{max},t) =0~\label{eq:B6}, 

\subsection{With Switching Rate Actuation}

We first consider Switching Rate actuated TCLs without the feature of the minimum dwell time. The pdf dynamic corresponding to a TCL unit with rate-switching can be described in this case by the PDE system
\begin{subequations} 
\begin{align} \label{eq:dyn_eps} 
  \frac{ \partial f_{0a} }{ \partial t}  &= - \frac{\partial}{\partial x} \bigg( u_0 f_{0a} \bigg) +  \frac{\partial^2}{\partial x^2} \left( \frac{1}{2} \sigma_0^2 f_{0a} \right) \\
  \frac{ \partial f_{0b} }{ \partial t} &= - \frac{\partial}{\partial x} \bigg( u_0 f_{0b} \bigg) +  \frac{\partial^2}{\partial x^2} \left( \frac{1}{2} \sigma_0^2 f_{0b} \right)  - \bar{\lambda}_1 f_{0b} + \bar{ \lambda}_0 f_{1b} \label{eq:dyn_eps_0b} \\
  \frac{ \partial f_{1b} }{ \partial t} &= - \frac{\partial}{\partial x} \bigg( u_1 f_{1b} \bigg) +  \frac{\partial^2}{\partial x^2} \left( \frac{1}{2} \sigma_1^2 f_{1b} \right)  + \bar{\lambda}_1 f_{0b} - \bar{\lambda}_0 f_{1b}  \label{eq:dyn_eps_1b} \\
  \frac{ \partial f_{1c} }{ \partial t} &= - \frac{\partial}{\partial x} \bigg( u_1 f_{1c} \bigg) +  \frac{\partial^2}{\partial x^2} \left( \frac{1}{2} \sigma_1^2 f_{1c} \right)  
\end{align}
\end{subequations}
together with the boundary conditions \eqref{eq:PDE_boundary}. The notation $\bar{\lambda}$ is used to extend the function $\lambda$ with zero values over the unsafe temperature distances $\Delta T_0$ and $\Delta T_1$.  In the case of a cooling unit, this translates into 
\begin{align}
 \bar{\lambda}_1 (\epsilon, T) &= \begin{cases}  
										0, ~T \in ( T_{\min}, T_{\min} + \Delta T_1 ) \\
                             						\lambda_1(\epsilon, T), ~T \in [ T_{\min}+ \Delta T_1, T_{\max} ) \\
                                                       \end{cases} \\
 \bar{\lambda}_0 (\epsilon, T) &= \begin{cases}  
										\lambda_0(\epsilon, T), ~T \in ( T_{\min} , T_{\max} - \Delta T_0 ] \\
										0, ~T \in ( T_{\max} -\Delta T_{0}, T_{\max} )~.
                                                       \end{cases} 
\end{align}

The reason why adding terms $\bar{\lambda}_1 f_{0b}$ and  $\bar{ \lambda}_0 f_{1b}$ gives a fitting dynamic in \eqref{eq:dyn_eps} is related to the exponential behavior of the survival and jump switch-rate times as $\Delta t \rightarrow 0$. We refer to \cite{bect2010unifying} for a more elaborate mathematical discussion in the context of GSHS. \\

We now include minimum dwell time conditions and consider the complete Switching Rate actuation. The idea is to continously track the part of the pdf that becomes locked for the external actuation. The same approach is used in \cite{zhang2013aggregated} and \cite{totu2014demand} in the context of the Switching Fraction actuation. We introduce two new density functions corresponding to the locked condition for mode ``off" and  for mode``on", $L_0:(-\infty, T_{\max}) \times [0,M_0) \times [0, \infty) \rightarrow \mathbb{R}_{+}$ and $L_1:(T_{\min}, \infty) \times [0,M_1) \times [0, \infty) \rightarrow \mathbb{R}_{+}$, defined as
\begin{align}
	L_i(x,y,t) = \frac{1}{dxdy}~ \mathrm{Pr}\big[ \hspace{3pt}T_t \in ( x, x + dx] \hspace{3pt} \wedge d_t \in  ( y, y+ dy] \hspace{3pt} \wedge  m_t=i  ~\big]~. \label{def:pdf_locked}
\end{align}
Using these pdfs, we can evaluate the part of $f_i(x,t)$ which remains responsive to the actuation. The terms $\bar{\lambda}_1 f_{0b}$ and  $\bar{ \lambda}_0 f_{1b}$  will thus be replaced in \eqref{eq:dyn_eps} by  $\bar{\lambda}_1 \left( f_{0b} - \int_{0}^{M_0} L_0(x,y,t) dy \right)$ and $\bar{\lambda}_0 \left( f_{1b} - \int_{0}^{M_1} L_1(x,y,t) dy \right)$.

The new pdfs must also be propagated in time. Their dynamic is given by normal Fokker-Planck equations, since no switching mechanism is active in the interior of the domains. These are
\begin{align}
   \frac{\partial L_{i}}{\partial t}(x,y,t) = - \frac{\partial}{\partial x} \bigg( u_i(x,t) L_{i}(x,y,t) & \bigg)   - \frac{\partial}{\partial y} L_{i}(x,y,t)  
    + \frac{\partial^2}{\partial x^2} \left( \frac{1}{2} \sigma_i^2(x,t) L_{i}(x,y,t) \right)~, \label{eq:FP_2D}
\end{align}
with boundary conditions that follow naturally,
\begin{subequations}  \label{eq:PDE_boundary_locked}  
\begin{align} 
    L_i(x,0,t) &= \bar{\lambda}_i \left(  f_{\bar{i}b} - \int_{0}^{M_{\bar{i}}} L_{\bar{i}} (x,y,t) dy \right) \label{eq:PBL_1} \\ 
    L_i(x,M_i,t) &= 0 \label{eq:PBL_2} \\
    L_0(-\infty,y,t) &= 0, ~L_0(T_{\max},y,t)=0 \label{eq:PBL_3}\\
    L_1(T_{\min},y,t)=0,~ L_1(\infty,y,t)=0 \label{eq:PBL_4}
\end{align}
\end{subequations}
where $i \in \{0,1\}$, and $\bar{i} =1-i$.  Eq. \eqref{eq:PBL_2}, \eqref{eq:PBL_3} and \eqref{eq:PBL_4} represent absorbing boundaries, while \eqref{eq:PBL_1} represents the incoming density current (or flow) of ``newly locked" for which the dwell time state $d_t$ has just been reseted to zero.

\section{Numerical simulation} \label{sec:numerical_sim}

This section verifies numerically the probability density model of the Switching Rate actuation, without the minimum dwell time feature. The verification procedure consists of two numerical simulations: a Monte Carlo analysis running multiple SHS model instances, and a finite dimensional linear approximation of the pdf PDE dynamics via a Finite Volume technique. The results show an equivalence between the two simulations.

The SHS simulation consists of time-discretized dynamics with a sample period $\tau_s=1$s. The SDEs are simulated with the Euler-Maruyama method.  The set-up is such that the control signal $\epsilon$ is constant during the sample period $\tau_s$, and rate-switches are generated as Bernoulli trials with success rate $1-e^{-\epsilon_i \tau_s}$, when the temperature is in the safe-zone. %An alternative way of generating rate-switches is to extract an arrival time according to the exponential distribution, and use it as countdown timer. A rate-switch event occurs when the timer runs out, and a new arrival time is extracted. If the temperature conditions or the external signal change in a relevant way before the timer has reached zero, a new arrival time is again extracted.

The PDE model is the basis for the second simulation. Eq. \eqref{eq:dyn_eps} together with boundary condition \eqref{eq:PDE_boundary} represent an infinite-dimensional dynamic. We approximate this dynamic with a finite-dimensional  form via the Finite Volume  Method (FVM), see e.g. \cite{ferziger2002computational}. This results in a numerical approximation of the weak solution of the PDE system. The FVM has the property of being locally and globally conservative, which will ensure that the probability in the system will always sum to one. We implement the FVM using an uniform grid, and a linear, cell-centered, piecewise-quadratic reconstruction scheme with an upstream flux rule. Because of Godunov's order barrier theorem, this third order accurate reconstruction scheme can create spurious oscillations, but no significant effects have been noticed in practice. Applying non-linear elements to the reconstruction scheme to correct this possibility is not an option, as it is important to obtain a dynamic that is linear in the state. We obtain a finite-dimensional dynamic of the following form,
\begin{align}
  \dot{F}_t &= (  A + B_0\epsilon_0  + B_1\epsilon_1 ) F_t,~ F_t \in \mathbb{R}^{n},~A, B_0, B_1 \in \mathbb{R}^{n \times n}. \label{eq:bilinear_form}
\end{align}

\begin{figure}[h!] 
\centering
\subfigure[Input signal A]{
      \includegraphics[width=0.47\textwidth]{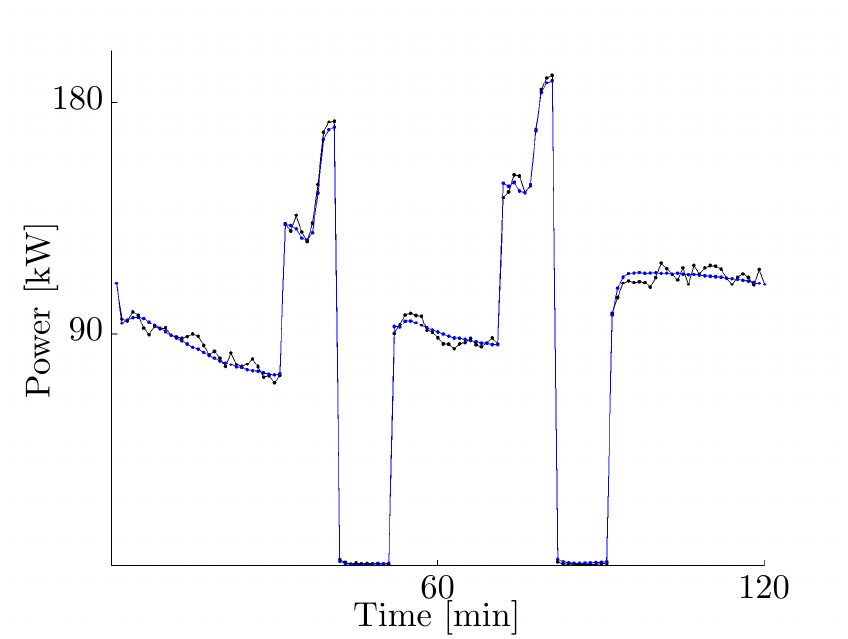}  \label{fig:P1}}  
\subfigure[Input signal B]{
      \includegraphics[width=0.47\textwidth]{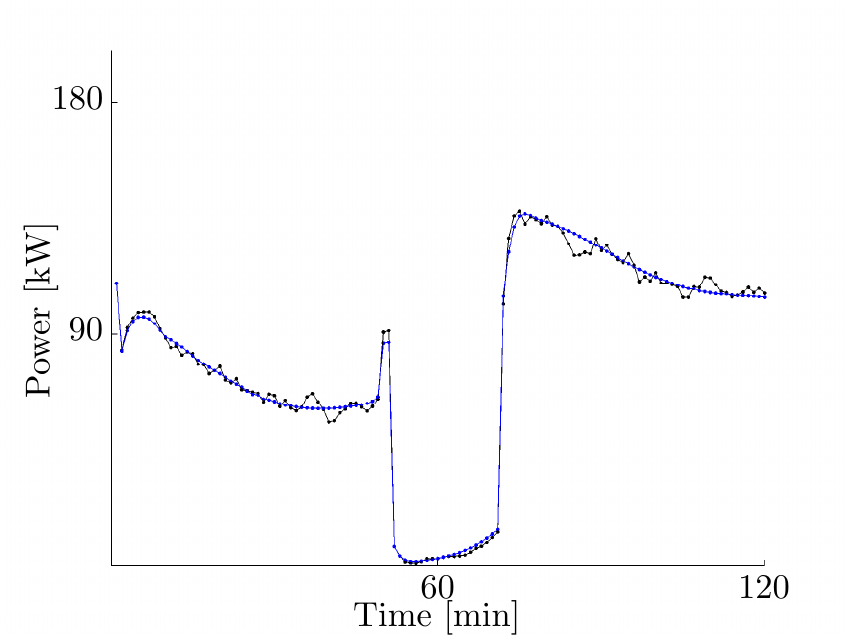}  \label{fig:P2}}  

\caption{Power consumption of the TCL population. The output of the Monte Carlo simulation is shown in black, and the PDE model is shown in blue.}
\label{fig:results_1}
\end{figure}

\begin{figure}[h!]
\centering
\subfigure[ Mode Off, signal A]{
      \includegraphics[width=0.47\textwidth]{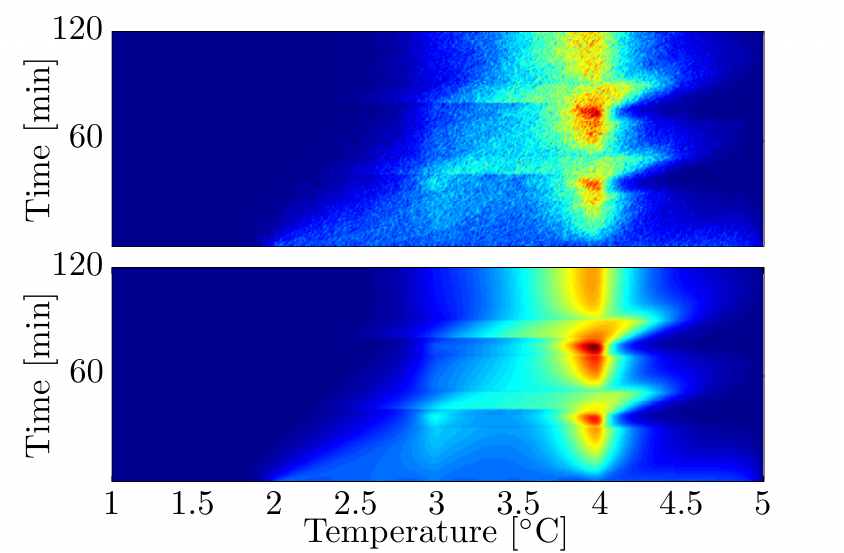}}  
\subfigure[ Mode On,  signal A]{
      \includegraphics[width=0.47\textwidth]{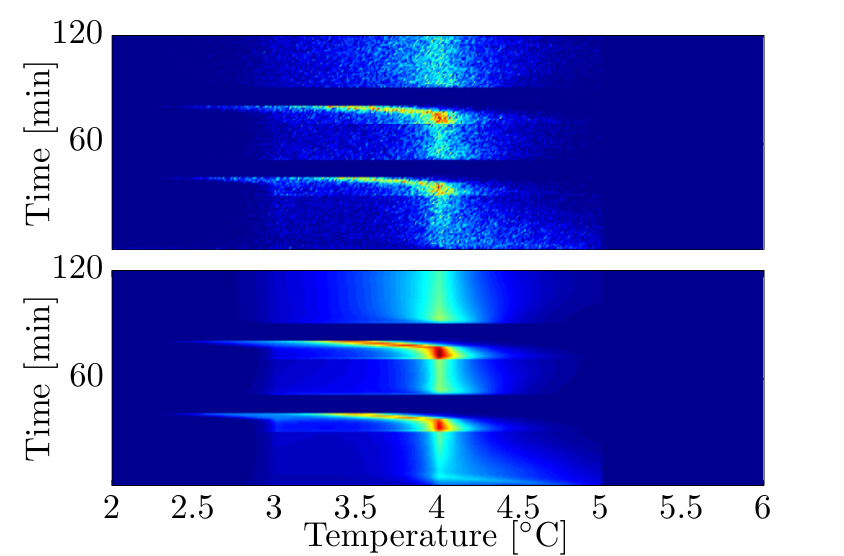}} 
\subfigure[ Mode Off, signal B]{
      \includegraphics[width=0.47\textwidth]{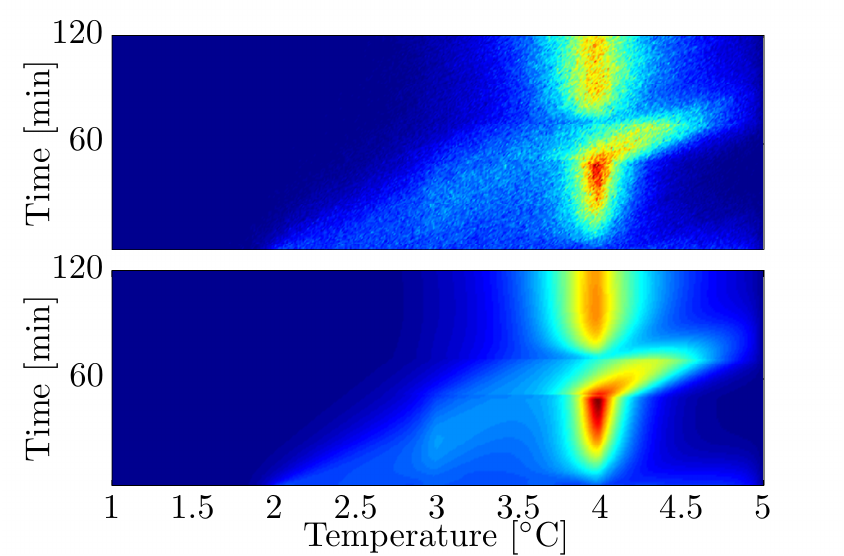}}
\subfigure[Mode ON, signal B]{
      \includegraphics[width=0.47\textwidth]{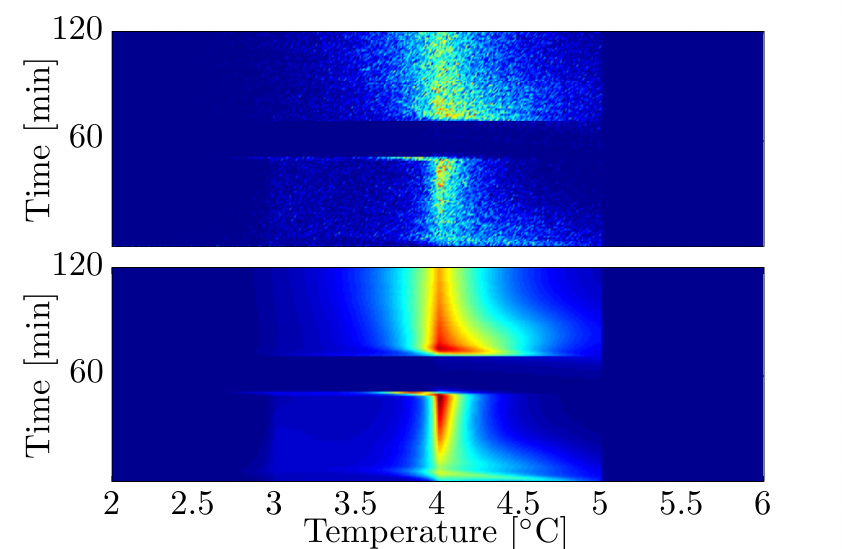}} 
\caption{Temperature densities across the TCL population. The empirical Monte Carlo pdf is shown in the top subplots, and the pdf from the PDE model is shown the bottom subplots. Blue represents low pdf values, and red high pdf values.}
\label{fig:results_2}
\end{figure}

We use the following TCL model elements $u_i(T,t)= aT +b_i$,  $\sigma_i(T,t) = \sigma$,  with parameter values $a=-1.5247^{-05}$, $b_0=3.6593^{-04}$, $b_1=-0.0026$, $\sigma=0.0065$, $T_{\min}=2$, $T_{\max}=5$, meant to approximate a refrigerator unit similar to \cite{totu2013control, totu2014demand}. Rate-functions $\lambda$ of the form \eqref{eq:simple_rate_functions} have been used. A practical deployment scenario requires a coordination center broadcasting the actuation signal $\epsilon_t=(\epsilon^0_t,\epsilon^1_t)$. Between broadcasts, the TCL units operate with the previously received values, resulting in a scenario with piecewise constant actuation. The broadcast sample period is $\tau_c=60$s. Simulations take place over a time horizon of  two hours, and two actuation signals are tested. These signals have a specially chosen form derived from the results in \cite{totu2014demand}, which is meant to show the power consumption flexibility. Figures \ref{fig:results_1} and \ref{fig:results_2} show comparisons between the Monte Carlo SHS simulation with 10000 identical units and the linear system model, for both pdf and power output. 

\section{Future Work} \label{sec:conclusion}
These succesful numerical results motivate future work, in two directions. First, the modeling result could be consolidated by more rigorous mathematical considerations, such as completing the GSHS description. Moreover, a two dimensional FVM scheme needs to be set up to introduce the minimum dwell time feature. Secondly, the bilinear model \eqref{eq:bilinear_form} can be analyzed for control. 


\bibliographystyle{eptcs}
\bibliography{MAIN}                 
\end{document}